# Do the unvaccinated disproportionately harm the vaccinated in a respiratory pandemic?

Denis G. Rancourt,* PhD, and Joseph Hickey, PhD

Correlation Research in the Public Interest
(correlation-canada.org)

* denis.rancourt@gmail.com

## Abstract

A parameter ψ was recently defined and introduced into the epidemiological modelling scientific literature, and is being accepted. The said parameter was used to argue that there was a disproportionate risk of infection incurred by vaccinated persons due to contacts with unvaccinated persons during the declared COVID-19 pandemic. Opposing published results show that, in general, there is virtually never a disproportionate risk to the vaccinated from the unvaccinated during a respiratory pandemic. Here, we show that the newly introduced vaccinology parameter ψ is incorrectly defined and that the conclusions of disproportionate risk are not valid. Specifically, we prove that the originating authors Fisman et al. (2022, 2024) incorrectly defined and applied the parameter ψ. Their application would imply that the said risk increases with increasing segregation from the unvaccinated (up to complete segregation), increases with increasing vaccination coverage (up to complete coverage) and increases with increasing vaccine efficacy (up to perfect vaccine efficacy), which is impossible. Use of the erroneous parameter ψ has the potential to encourage unnecessarily aggressive public health policies and interventions.

**Table of contents**

## 1. Introduction

Simple mathematical models are routinely used to simulate scenarios of infectious disease dynamics (Keeling and Rohani, 2008). Such models are extensively applied to predict outcomes and to guide policy decisions regarding imposed measures, as was the case during the declared COVID-19 pandemic.

A common model type is to use the SIR approach, in which a susceptible person (S) comes into contact with an infectious person (I), the susceptible person can become infectious, and infectious people eventually recover (R) and become permanently immune. Vaccination can be added. The model produces the temporal evolution of the three groups (S, I and R) in a given hypothetical society of thus interacting individuals.

In the models of interest in the present paper, SIR models of epidemic dynamics were implemented with two interacting societal groups (vaccinated and unvaccinated) to examine epidemic outcomes for variable degrees of interaction between the two groups, including whether the unvaccinated put the vaccinated unduly or disproportionately at risk, using epidemiological parameters intended to be representative of severe acute respiratory syndrome coronavirus 2 (SARS-CoV-2). The said epidemiological parameters are typical of those used for most viral respiratory diseases.

The limitations, of course, are that the model assumptions and parameters must approximate the actual disease transmission and inter-personal contacts occurring in the real society.

Fisman et al. (2022) and Fisman et al. (2024) (same co-authors, with the addition of Simmons in 2024) applied such an SIR model (vaccinated + unvaccinated) but unnecessarily introduced an *ad hoc* parameter ψ, which they used to conclude that there was a disproportionate risk of infection incurred by vaccinated persons due to contacts with unvaccinated persons during the COVID-19 pandemic. We made a space-limited open comment to their first paper (Fisman et al., 2022), on the website of the publishing journal (Rancourt and Hickey, 2022). Their articles are now amply cited but their unusual (and incorrect) method and surprising conclusion have not been challenged in a scientific paper.

On the other hand, Hickey and Rancourt (2023a) made a broad-parameter-range exploration of a generalized SIR, vaccinated + unvaccinated model and found that the risk of infection incurred by vaccinated persons due to contacts with unvaccinated persons is virtually never disproportionate, including for the parameters used by Fisman et al. (2022, 2024).

Hickey and Rancourt (2023b) found similar results with a generalized SIR, robust + vulnerable model to conclude, from a purely infectious transmission perspective, that: "Isolated care homes of vulnerable residents are predicted to be the worst possible mixing circumstances for reducing harm in epidemic or pandemic conditions.".

In this paper, we provide an analytic proof that the newly introduced vaccinology parameter ψ is incorrectly defined and that the conclusion of disproportionate risk from the unvaccinated is not valid. This is important because an incorrect conclusion can lead to harmful public health policies.

## 2. Nature of the error made by Fisman et al.

Fisman et al. (2022, 2024) modelled hypothetical COVID-19 epidemics, assuming interacting groups of vaccinated and unvaccinated people, segregated to different degrees η, between 0 and 1, where η = 1 corresponds to total segregation (zero contacts between vaccinated and unvaccinated people).

In their model, Fisman et al. (2022, 2024) have introduced a new parameter (ψ), which (to our knowledge, having examined the literature), has not been used or adopted by other research groups that perform modelling.

Their mathematical definition of ψ can be proven to be incorrect on three grounds, which are disjunctive and individually sufficient:

(1) ψ in its mathematical definition does not represent what Fisman et al. state it to represent
(2) ψ — while stated to represent the physical reality of risk incurred by vaccinated persons due to contacts with unvaccinated persons — diverges to exceedingly large values in approaching model parameter bounds

(3) ψ — while stated to represent the physical reality of risk incurred by vaccinated persons due to contacts with unvaccinated persons — varies with changing model parameters in directions contrary to reason

We prove these points (1, 2 and 3) below. See Sections 5.1, 5.2 and 5.3, and also Section 6.

Fisman et al. (2022, 2024) explicitly rely on ψ for main conclusions in both their papers. They base these conclusions in their two papers solely on ψ. The conclusions do not follow from ψ values, and are incorrect. The unfounded conclusions, if propagated, constitute a potential source of harm via thus misinformed public health policy and interventions.

The present refutation of the Fisman et al. (2022, 2024) errors related to their introduced ψ parameter is as succinct as possible, but it is more intricate than many readers might want. This is unavoidable because of the so-called bullshit asymmetry principle, also known as Brandolini's law: *"The amount of energy needed to refute bullshit is an order of magnitude bigger than to produce it."* (Gelman, 2019).

## 3. How do Fisman et al. define the parameter ψ?

In the first paper, Fisman et al. (2022) define ψ as:

> "the fraction of all infections among vaccinated people that derived from contact with unvaccinated people, divided by the fraction of all contacts [involving vaccinated people] that occurred with unvaccinated people".

For convenience, we write this as:

$$\psi = B_v / f_{vu} = (H_{vu} / H_v) / (C_{vu} / C_v) \qquad (1)$$

where

- $B_v$ is the fraction of all infections among vaccinated people that derived from contact with unvaccinated people (over the duration of the considered time period)

- $f_{vu}$ is the fraction of all contacts involving vaccinated people that occurred with unvaccinated people
- $H_{vu}$ is the number of infections among vaccinated people that derived from contact with unvaccinated people (over the duration of the considered time period)
- $H_v$ is the number of all infections among vaccinated people (over the duration of the considered time period)
- $C_{vu}$ is the number of contacts involving vaccinated people that occurred with unvaccinated people (over the duration of the considered time period)
- $C_v$ is the number of all contacts involving vaccinated people (over the duration of the considered time period)

We re-coded the model (e.g., Hickey and Rancourt, 2023a) and established that this is indeed exactly the definition that Fisman et al. (2022) applied in producing their figures.

In the second paper, Fisman et al. (2024) state that they are using the same ψ parameter: "As in our earlier work we estimate a quantity that we denote ψ …". Note that there is again no cited reference, and that such a parameter has not, to our knowledge, previously been introduced in epidemiology or vaccinology, nor has introduction of such a parameter been motivated (e.g., Chapter 8 in Keeling and Rohani, 2008; Bjørnstad et al., 2020).

Fisman et al. (2024) then follow the above phrase with the only explicit definition of ψ stated in the text of the body of their 2024 paper, as:

> "As in our earlier work we estimate a quantity that we denote ψ, defined as the incidence of infections among the vaccinated population derived from contact with unvaccinated people, divided by the fraction of the population that is unvaccinated."

This definition of ψ is unambiguously different from the definition expressed in Equation 1 (above), and would be written as:

$$\psi = B_v / (1 - P_v) \qquad (2)$$

where $P_v$ is the fraction (0 to 1) of the population that is vaccinated.

However, the definition of ψ given in the figure caption of Fisman et al. (2024)'s Fig. 3 is:

> "ψ is the ratio of the fraction infections acquired by vaccinated people from unvaccinated people divided by fraction of contacts with unvaccinated people." (*sic*)

Typo notwithstanding, this is the same as Equation 1 (above), the same as introduced by Fisman et al. (2022) in the first paper.

We verified Fisman et al. (2024)'s code, provided by them as an Excel calculation (their reference 26), and established that the two papers apply exactly the same definition of ψ: Equation 1 (above), and not the definition given in the text of the body of their 2024 paper (not Equation 2, above).

## 4. How do Fisman et al. interpret the parameter ψ?

Main interpretive statements regarding ψ, in Fisman et al. (2022, 2024) are:

1. "As like-with-like mixing increased (i.e., with reduced contact between vaccinated and unvaccinated subpopulations) … the contribution of risk to vaccinated people caused by infection acquired from contact with unvaccinated people (as measured by ψ) increased. The larger the value of ψ, the more unvaccinated people contributed to infections in the vaccinated subpopulation." (2022: at pg. E575)

2. "The quantity ψ … remained above 1 throughout, signifying a disproportionate contribution to infection risk to vaccinated people by the unvaccinated population. When we estimated ψ cumulatively, the value at 10 years was approximately 2.14, meaning that infection among vaccinated people was more than twice as likely to have been acquired from unvaccinated people than would have been expected based on contact rates alone." (2024: at pg. 5/13)

3. "(Fig 4). Cumulative ψ rose as like-with-like mixing increased, but was elevated across all scenarios, indicating disproportionate contribution to risk among the vaccinated from the unvaccinated group." (2024: pg. 7/13)

4. "… while the contact-adjusted risk to vaccinated individuals associated with contact with unvaccinated groups is disproportionate." (2024: pg. 10/13, last sentence of the article)

The above statements by Fisman et al., while not being optimally clear and constant, nonetheless allow one to infer that they believe ψ to measure:

- "the contribution of risk to vaccinated people caused by infection acquired from contact with unvaccinated people"

or

- the likelihood that "infection among vaccinated people was … acquired from unvaccinated people" compared to what "would have been expected based on contact rates alone"

In addition, Fisman et al. appear to imply that one can use ψ to ascertain whether the said risk or likelihood is "disproportionate", and that ψ measures the degree to which the risk or likelihood is "disproportionate". This is addressed in Section 7, below.

## 5. Proof that the Fisman et al. definition and interpretation of the parameter ψ are incorrect

### 5.1   ψ does not represent what Fisman et al. state it to represent

The statement "the contribution of risk to vaccinated people caused by infection acquired from contact with unvaccinated people (as measured by ψ)" means that ψ is a (valid and useful) measure of "the contribution of risk to vaccinated people caused by infection acquired from contact with unvaccinated people". This is incorrect.

The quantity $B_v$ [the fraction of all infections occurring in vaccinated people caused by contacts with unvaccinated people, $B_v = H_{vu} / (H_{vu} + H_{vv}) = H_{vu} / H_v$] is exactly "the contribution of risk to vaccinated people caused by infection acquired from contact with unvaccinated people". It measures the fraction of infections in the vaccinated, which were caused by contact with the unvaccinated. One cannot have a more direct measure of the risk in question.

Instead of interpreting and reporting $B_v$, Fisman et al. (2022, 2024) introduced the new parameter ψ, in which $B_v$ is divided by $f_{vu}$ [the fraction of all contacts experienced by vaccinated people, irrespective of whether the contacts were infectious or not, which were with unvaccinated individuals, $f_{vu} = C_{vu} / (C_{vu} + C_{vv}) = C_{vu} / C_v$].

It is not clear why one would define a new parameter in this way, as ψ = $B_v / f_{vu}$ (Equation 1, above), since $B_v$ itself is already a dimensionless ratio, the desired probability, bounded between 0 and 1, resulting from the full societal structure (η, $P_v$, etc.) and the contact frequencies. $B_v$ does not need to be normalized or adjusted.

Fisman et al. appear to be under the incorrect impression that their definition of ψ provides the benefit of a "contact-adjusted risk to vaccinated individuals associated with contact with unvaccinated groups". Contact normalization of risk would involve putting the risk of infection on a per contact basis. ψ does not achieve this. Likewise, ψ is not a relative risk, as it is not a ratio of risks of the same type.

This proves that ψ in its mathematical definition does not represent what Fisman et al. state it to represent. Next, we examine how exceedingly large values and unreasonable variations of ψ result from its definition.

## 5.2   ψ diverges to exceedingly large values

Examining the newly introduced *ad hoc* parameter ψ further, it can be written:

$$\psi = B_v / f_{vu} = B_v / (C_{vu} / C_v) = B_v (C_v / C_{vu}) \qquad (3)$$

And — given the definitions of η and $P_v$ — an exact expression for $f_{vu}$ is:

$$f_{vu} = C_{vu} / C_v = (1 - \eta)(1 - P_v) \quad (4)$$

Giving:

$$\psi = B_v / (1 - \eta)(1 - P_v) \quad (5)$$

In other words, in defining ψ, Fisman et al. multiply the correct and exact expression for "the contribution of risk to vaccinated people caused by infection acquired from contact with unvaccinated people" ($B_v$, which is bounded between 0 and 1) by the unrelated quantity $1/f_{vu}$ (or $C_v/C_{vu}$), which necessarily (by its definition, and in view of Equations 4 and 5) is always greater than 1 and mathematically tends to infinity (+∞) as η → 1 (η = 1 corresponds to complete segregation between the vaccinated and unvaccinated groups, or complete so-called like-with-like mixing) or as $P_v$ → 1 ($P_v$ = 1 corresponds to the entire population being vaccinated, no unvaccinated individuals). This is why Fisman et al. incorrectly report much disproportionality, which they take to follow from ψ > 1 (see Section 7, below).

To formalize this, and to remove any ambiguity:
- as η → 1 or as $P_v$ → 1, $C_{vu}$ (the vaccinated-person contacts from unvaccinated persons) must go to zero
- as η → 1 or as $P_v$ → 1, the value of $C_v$ is a finite non-zero positive number since it is simply the number of contacts of any type experienced by the vaccinated group (it is zero only if $P_v$ = 0)
- therefore, as η → 1 or as $P_v$ → 1, $1/f_{vu} = C_v/C_{vu}$ tends to infinity (which equivalently also follows from Equation 4)

This provides a mathematical mechanism whereby ψ can acquire exceedingly large values.

We show by calculations illustrated in the figures below that this does indeed occur. The parameter ψ — by virtue of its incorrect design or definition — exhibits exceedingly large values that cannot reasonably be assigned to the actual risk said to be of concern.

### 5.3   ψ varies in directions contrary to reason

By virtue of its definition, $1/f_{vu}$ = $C_v/C_{vu}$ = $1/((1−η)(1−P_v))$ not only is always greater than 1, and diverges to infinity as η → 1 or as $P_v$ → 1, but also always increases with increasing η or with increasing $P_v$.

The figures below show that ψ also increases with increasing η (**Figure 1**) or with increasing $P_v$ (**Figure 2**), almost always. In addition, ψ often increases with increasing vaccine efficacy VE (**Figure 1**, **Figure 3**).

These behaviours of ψ (shown in the figures below) would imply — adopting Fisman et al.'s incorrect interpretation of the meaning of ψ — that the "infection risk to vaccinated people by the unvaccinated population" or the "contact-adjusted risk to vaccinated individuals associated with contact with unvaccinated groups" *increases* (often strongly) as one:

- increases segregation between the vaccinated and unvaccinated groups (η); and/or
- increases the share of the population that is vaccinated ($P_v$); and/or
- increases vaccine efficacy (VE)

These are absurd consequences resulting from the definition and interpretation of ψ, the opposite of what must occur as η, $P_v$ and VE are increased. The solution to this paradox is that the parameter ψ is an *ad hoc* parameter that is ill-defined and that does not mean what Fisman et al. take it to mean.

## 6. Demonstration of the unrealistic behaviour of the parameter ψ

As noted above, $B_v$ is directly the risk of an infected vaccinated person having acquired their infection from an unvaccinated person. The said risk can then be evaluated as a function of the segregation of the two groups (η), vaccine efficacy (*VE*) or population fraction of vaccinated (*$P_v$*), and so on. In each case *$B_v$* has a straightforward interpretation of interest to policy makers (e.g., Hickey and Rancourt, 2023a, 2023b).

Therefore, it is useful to compare graphs of ψ with graphs of *$B_v$*, versus key model parameters, which is done in the following. This quantitatively illustrates the anomalous behaviours of ψ.

In particular, the Fisman et al. (2022) statement that "The larger the value of ψ, the more unvaccinated people contributed to infections in the vaccinated subpopulation" is mathematically incorrect. The opposite is generally true, as shown below.

Figures 1-3 show ψ (left panels) and *$B_v$* (right panels) versus η, *$P_v$* and *VE*, respectively.

All graphs in Figures 1-3 use the model parameters considered by Fisman et al. (2022): β = 437 $yr^{-1}$ (probability of transmission per contact multiplied by number of contacts per year), γ = 73 $yr^{-1}$, *NI* = 0.2, and *$P_v$* = 0.8 if not otherwise indicated.

Here, *NI* is the proportion of the unvaccinated population that has pre-existing natural immunity at the outset of the epidemic. In the second paper Fisman et al. (2024) do not admit any pre-existing natural immunity, and do not make any mention of "natural immunity".

In Figures 1a and 1b (which have expanded Y-scales compared to Figures 1c and 1d), we note that *$B_v$* decreases with increasing η, whereas ψ increases with increasing η (for *VE* > *NI*). According to Fisman et al., this behaviour of ψ would mean that "the contribution of risk to vaccinated people caused by infection acquired from contact with unvaccinated people" increases with increasing like-with-like mixing (i.e., with increasing segregation between vaccinated and unvaccinated), and is largest for complete segregation (η = 1), even though *$B_v$* itself decreases to zero at η = 1 (for all values of *VE*, up to 1, see Figures 1c and 1d and the inset of Figure 1c), which is impossible.

Fisman et al.'s ill-defined "risk from unvaccinated people" (ψ) that monotonically goes to its largest value when the agent of risk is removed (η = 1) is an absurdity, and is therefore of no valid utility.

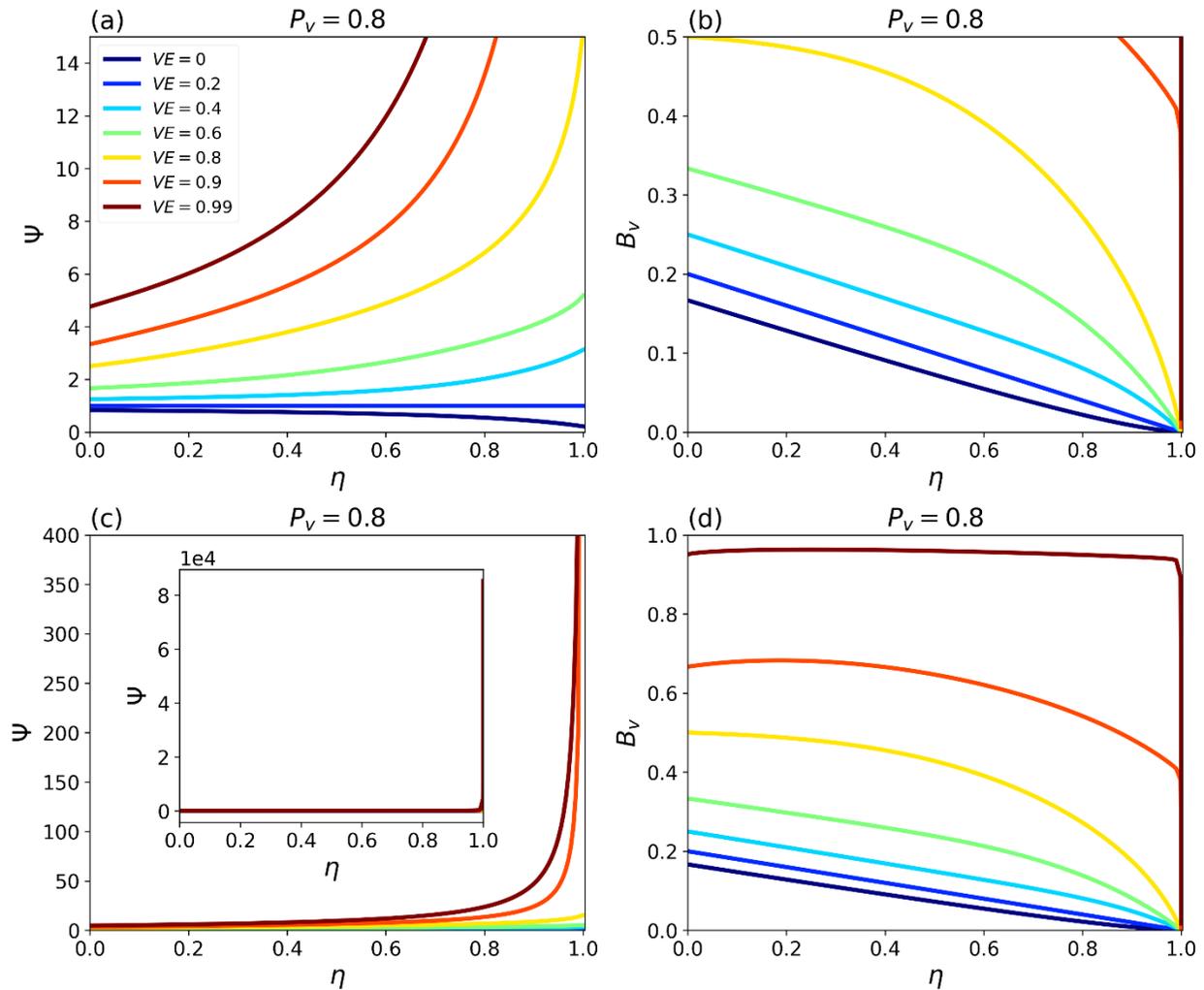

Figure 1: Panels (a) and (c): Fisman et al.'s ψ index vs the degree of segregation η, for different values of the vaccine efficacy *VE*. Panels (b) and (d): the fraction of all infections among vaccinated people that derived from contact with unvaccinated people, $B_v$, vs η, for different values of *VE*. Panels (a), (b), and (c) have expanded Y-axes, and the inset of panel (c) shows the full span of the Y-axis for the ψ vs η curves.

The meaningless result of largest "risk from unvaccinated people" ($\psi$) when nothing happens ($\eta$ = 1), and increasingly so for increasing *VE* (when the vaccinated should be increasingly protected), is maintained and amplified for larger values of *VE*, as shown in Figure 1c. Here, Figure 1 illustrates the behaviours as $\eta \rightarrow 1$ (using $\eta$ = 0.999, 0.9999, …). Note that, although it is not apparent in the figure, $\psi$ itself (as opposed to its factor $1/f_{vu}$) does not tend to infinity as $\eta \rightarrow 1$. Rather, it converges to a large finite value for each value of VE.

The curves with *VE* > *NI* (*VE* = 0.4 through 0.99) in **Figure 1** prove that the Fisman et al. (2022) statement "The larger the value of $\psi$, the more unvaccinated people contributed to infections in the vaccinated subpopulation" is unambiguously incorrect. We see that the opposite is true: on increasing segregation $\eta$, where $\psi$ increases (left panels), $B_v$ decreases (right panels). $B_v$ is defined "as the share of infections among vaccinated people that were due to contacts with infectious unvaccinated people", which is exactly the proportion that "unvaccinated people contributed to infections in the vaccinated subpopulation".

Here (**Figure 1**), as the vaccine is made perfectly effective (near-1 values of *VE*), "the contribution of risk to vaccinated people caused by infection acquired from contact with unvaccinated people", stated incorrectly by Fisman et al. to be represented by the parameter $\psi$, now diverges ($\psi$ diverges) to large values, orders of magnitude outside of any reasonable range compared to its values at the lower values of $\eta$ (Figure 1c).

On the contrary, for *VE* near 1, the actual risk $B_v$ does not diverge as one approaches $\eta$ = 1, and there is virtually no change in $B_v$ versus segregation $\eta$, up to nearly complete segregation (Figures 1b and 1d). It is difficult to reconcile the Fisman et al. "disproportionate" risk from the unvaccinated (inferred from the parameter $\psi$ as interpreted by Fisman et al. (2022, 2024): "their choices affect risk of viral infection among those who are vaccinated in a manner that is disproportionate to the portion of unvaccinated people in the population") with $B_v$ decreasing with increasing $\eta$ (**Figure 1**), while being virtually independent of $\eta$ for large *VE* (**Figure 1**). In plain terms: In the natural world, separation from the cause of risk reduces the risk. In the natural world, there is no risk to the vaccinated, "disproportionate" or other, from the unvaccinated when $VE$ = 1.

**Figure 1** proves that the main stated inference of Fisman et al. based on the parameter ψ ("disproportionate" to the degree that ψ is large) is incorrect.

This should be sufficient to convince readers of the incorrect approach of introducing and thus interpreting the parameter ψ, but there is more.

**Figure 2** shows ψ and $B_v$ versus $P_v$, for many values of η. Here, $B_v$ monotonically decreases to zero as $P_v$ is increased to 1, as expected, and has smaller and smaller values for larger and larger values of η.

In contrast, the "contribution of risk to vaccinated people caused by infection acquired from contact with unvaccinated people", as incorrectly stated by Fisman et al. to be represented by the parameter ψ, monotonically increases as $P_v$ is increased, and is systematically larger for larger values of η. This would mean that the "contribution of risk to vaccinated people caused by infection acquired from contact with unvaccinated people" increases the more people are vaccinated, and the more the vaccinated are segregated from the unvaccinated, up to the extreme values of total vaccination coverage ($P_v$ = 1), and up to total segregation (η = 1), and both total vaccination coverage and total segregation ($P_v$ = 1 and η = 1), which is impossible.

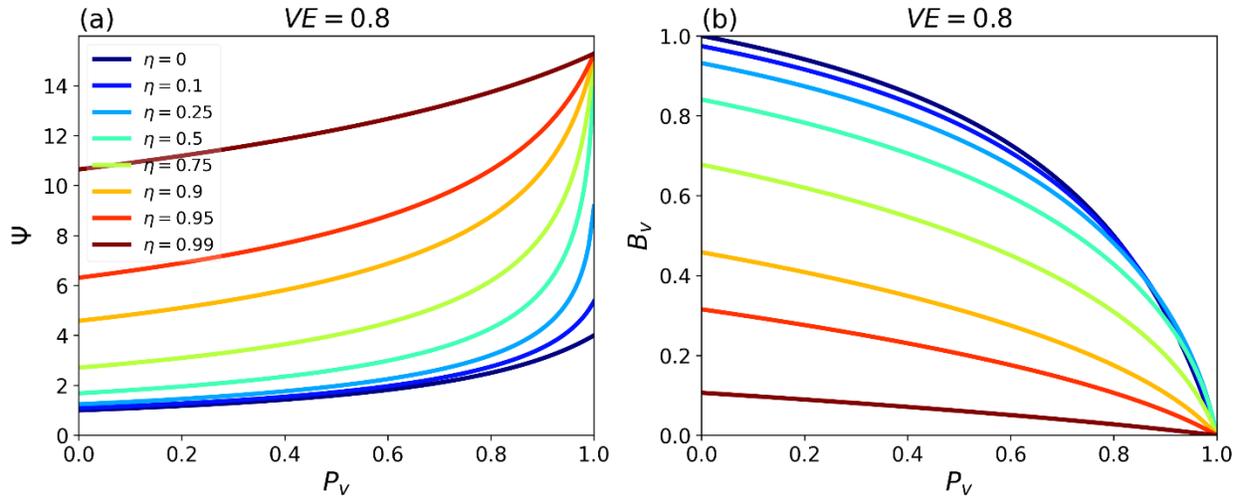

**Figure 2:** (a) ψ vs the population fraction of vaccinated people $P_v$ for different values of the degree of segregation η. (b) $B_v$ vs $P_v$ for different values of η.

We learn, for example (Figure 2a), that the supposed "contribution of risk to vaccinated people caused by infection acquired from contact with unvaccinated people" or likelihood that "infection among vaccinated people was … acquired from unvaccinated people" compared to what "would have been expected based on contact rates alone" (ψ) is approximately "15" at total vaccine coverage ($P_v$ = 1) and for virtually total segregation (η → 1) when *VE* = 0.8, which would be a stunning result if it were not meaningless.

In all the ψ-$B_v$ pairs of curves shown in the two panels of **Figure 2**, for each of the values of η shown, the larger the value of ψ, the smaller the value of $B_v$ (the less unvaccinated people contributed to infections in the vaccinated subpopulation); which again proves that the Fisman et al. (2022) statement "The larger the value of ψ, the more unvaccinated people contributed to infections in the vaccinated subpopulation" is rigorously incorrect.

In every case, we demonstrate that their *ad hoc* parameter ψ is not what Fisman et al. believe it to be.

Finally, **Figure 3** shows ψ and $B_v$ versus vaccine efficacy *VE*, for many values of segregation η. Here, $B_v$ goes to 1 in the limit as *VE* approaches 1, trivially because all of the infections of vaccinated individuals must come from unvaccinated individuals as *VE* approaches 1. Nonetheless, $B_v$ values are systematically smaller for larger and larger segregation η, as expected; and $B_v$ is essentially independent of η when *VE* is near 1, as also shown above (**Figure 1**).

The opposite is true of the parameter ψ (Figure 3a) and its interpretation made by Fisman et al.: the greater the segregation η, the more ψ increases; to astronomical values, as one goes to the larger values of *VE*. This inversion (compared to the behaviour of $B_v$) would mean that the more one is segregated, the greater the danger from being infected by the individuals one is segregated from, and all the more so if *VE* is high. This is difficult to reconcile with reality. In fact, it is nonsense arising from the ill-defined "contact-adjustment" used by Fisman et al. to construct the parameter ψ (i.e., division by $f_{vu}$, see Equations 1, 3, 4 and 5).

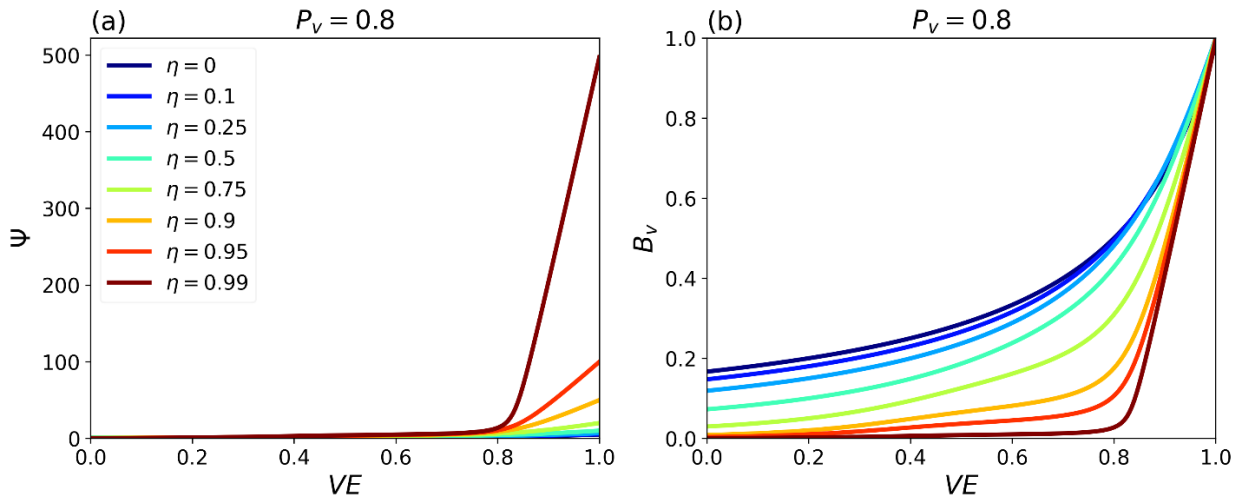

**Figure 3: (a) ψ vs the vaccine efficacy *VE*, for different values of the degree of segregation η. (b) $B_v$ vs *VE* for different values of η.**

The fundamental problems with and absurdities arising from the parameter ψ are not removed in the second paper by Fisman et al. (2024), by adding "diminished vaccine efficacy for preventing infection with the emergence of Omicron SARS-CoV-2 variants, waning immunity, the impact of prior immune experience on infectivity, 'hybrid' effects of infection in previously vaccinated individuals, and booster vaccination" or by evaluations "over a 10-year time horizon".

For example, the following approximate values of cumulative ψ at time 0.2 years (after the epidemic wave) occur using the case featured by Fisman et al. (2024) ("Value" column in their Table 1), with differing values of segregation η and vaccine efficacy VE (Table 1, below).

Table 1: Cumulative ψ at 0.2 years, Fisman et al. (2024) featured case, with differing values of η and VE, calculated using their Excel calculator (their reference 26).

| η (segregation) | VE (vaccine efficacy) | cumulative ψ at 0.2 years |
| --- | --- | --- |
| 0.95 | 0.8 | 8.5 |
| 0.99 | 0.99 | 160 |
| 0.999 | 0.99 | 1500 |

Following the erroneous interpretation of Fisman et al. (2024), these values of cumulative ψ would mean "that infection among vaccinated people [after 0.2 years] was [8.5 times, 160 times, 1500 times…] as likely to have been acquired from unvaccinated people than would have been expected based on contact rates alone", which is impossible, at these values of segregation η and vaccine efficacy VE (Table 1).

Following the erroneous interpretation of Fisman et al. (2024), the supposed risk would paradoxically increase with increasing segregation η and increasing vaccine efficacy VE (Table 1, above). Again: In the natural world, separation from the cause of risk (here, the unvaccinated) reduces the risk. In the natural world, there is no risk to the vaccinated, disproportionate or other, from the unvaccinated when η = 1 or when VE = 1. The Fisman et al. interpretation of the parameter ψ is contrary to what is possible in reality.

# 7. Is the risk to the vaccinated from the unvaccinated ever actually disproportionate?

The authors who introduced the new parameter ψ into the scientific literature, Fisman et al., use the word "disproportionate" or "disproportionately" regarding the risk from the unvaccinated 5 times in the first paper (2022) and 11 times in the second paper (2024). What do Fisman et al. mean, exactly?

It is not clear what Fisman et al. mean by disproportionate risk, beyond meaning that the considered value of their newly introduced parameter ψ is larger or much larger than 1.

In the Conclusion section of the first paper, Fisman et al. (2022) make a relatively definitive statement about disproportionality, incompatible with other statements they make involving the parameter ψ, as:

> "Using simple mathematical modelling, we have shown that, although risk associated with avoiding vaccination during a virulent pandemic accrues chiefly to those who are unvaccinated, the choice of some individuals to refuse vaccination is likely to affect the health and safety of vaccinated people in a manner disproportionate to the fraction of unvaccinated people in the population."

Well, the "likely[hood] to affect the health and safety of vaccinated people" is exactly the parameter $B_v$ (= $H_{vu}/H_v$) ("the fraction of all infections among vaccinated people that derived from contact with unvaccinated people"), and "the fraction of unvaccinated people in the population" is $1 - P_v$.

Therefore, adopting this particular explanation by Fisman et al. (2022) quoted above, disproportionality would occur when $B_v > (1 - P_v)$, or when $B_v/(1 - P_v) > 1$, which has nothing to do with the parameter ψ.

Actually, $B_v/(1 - P_v)$ is often moderately greater than 1, for moderate segregation η. This is illustrated, for example with VE = 0.8, in **Figure 2**, above. Here, $B_v = (1 - P_v)$ would be the straight line extending from $(P_v, B_v) = (0, 1)$ to $(P_v, B_v) = (1, 0)$.

Contrary to the incorrect interpretation advanced and implied by Fisman et al. (2022, 2024) — that the degree of disproportionality is measured by their parameter ψ — any disproportionality or "underproportionality" ($B_v/(1 - P_v)$) of risk to the vaccinated from the unvaccinated is always moderate (no large or divergent values occur), larger than 1 for small η, and smaller than 1 for large η. Contrary to the interpretations of Fisman et al., the vaccinated are the greatest risk to each other when they have little or no contacts with the unvaccinated, which is an unavoidable consequence of the segregation.

In general: The actual disproportionality ($B_v/(1 - P_v)$) is a trivial consequence of vaccination and mixing. It is never as large as the large values of the *ad hoc* parameter ψ, and it can be greater, smaller, or equal to 1, depending on the parameters η, $P_v$, VE, NI, waning times, etc.

Fisman et al. (2022, 2024) and the citing authors who follow them cannot have it both ways, simultaneously using both an incorrect and a correct meaning of "disproportionate": mostly the parameter ψ, and a spattering of $B_v/(1 - P_v)$. See also the discussion of Equations 1 and 2 above, arising from contradictory statements in the second article (Fisman et al., 2024).

## 8. Conclusion: There is an obligation not to apply the parameter ψ

As demonstrated herein, Fisman et al. (2022, 2024) define and apply an *ad hoc* parameter ψ in such a way as to imply that the risk of infection to the vaccinated from the unvaccinated is disproportionate, to a degree ψ, which is artificially and unrealistically large.

The resulting incorrectly inferred risk is unrelated to the actual risk, is virtually always larger than the actual risk, can be unrealistically large, and it nonsensically increases on increasing segregation η and/or increasing vaccinated population fraction $P_v$ and/or increasing vaccine efficacy VE.

The ψ-based interpretation advanced by Fisman et al. (2022, 2024) has a potential to cause harm by encouraging overly aggressive public health policies of intervention.

Therefore, the error of Fisman et al. should be corrected with publishers' notices, and in the citing literature, or it will make its way into established venues.

For example, a recent cross-sectional survey cites Fisman et al. (2022) as (Are et al., 2024, their reference 15):

> "Modelling studies have explored the impact of homophily in a range of contexts, and its impact on transmission dynamics is well documented [13, 14]. For instance, one modelling study argued that the mixing of vaccinated and unvaccinated groups contributes to a considerable risk of infection for the vaccinated group, occurring at a rate that is disproportionately higher than what would be expected based solely on the contact between the two groups [15]."

The review of Are et al. (2024) is, in turn, cited by the review of González-Parra et al. (2024) entitled "Learning from the COVID-19 pandemic: A systematic review of mathematical vaccine prioritization models", and so on.

The first article by Fisman et al. (2022) has, to date, been cited 65 times, according to Google Scholar. This shows that the erroneous methods and interpretations of Fisman et al. (2022) are diffusing into the scientific literature with no visible opposition.

As another example, one that is directly political, we have the following 29 April 2022 statements by lawmakers in Canada's Parliament referring to and relying on the first paper of Fisman et al. (2022) (Hansard-61, 2022, at pp. 4545-4546, emphasis added):

**Mr. Blaine Calkins (Red Deer—Lacombe, CPC):**

Madam Speaker, if the NDP-Liberals will not follow the province's lead and give unvaccinated Canadians their rights back, maybe they will follow our international partners. We know that the Prime Minister values his playboy image on the world stage more than anything else, as his travels and selfies prove, but our international partners are bewildered as to why the Canadian government is so reluctant to let life return to normal for all Canadians.

Switzerland and Greece are removing all travel-related restrictions next week and virtually no other country requires them for domestic travel for their citizens, *so why will the government not follow the science*?

**Mr. Adam van Koeverden (Parliamentary Secretary to the Minister of Health and to the Minister of Sport, Lib.):**

Madam Speaker, I thank my hon. colleague for giving me the opportunity to *highlight a recent study indicating that unfortunately the unvaccinated continue to disproportionately risk the safety of those vaccinated against COVID-19*, and highlight the fact that in the last week, almost 12,000 Canadians have received their first dose, 30,000 Canadians have received their second, 89,000 received their third and over 220,000 have received their fourth dose.

This pandemic is not over. We all want it to be, but *we must continue to follow the science*, we must continue to support our neighbours, we must continue to fight for kids to ensure that young people under six do not get COVID-19, as they have been the biggest numbers in the last couple of days.

This demonstrates in concrete terms that bad science can influence policy and law making.

# References

Are et al. (2024): Are, E.B., Card, K.G. & Colijn, C. /// The role of vaccine status homophily in the COVID-19 pandemic: a cross-sectional survey with modelling /// *BMC Public Health* 24, 472 (2024). https://doi.org/10.1186/s12889-024-17957-5

Bjørnstad et al. (2020): Bjørnstad, O.N., Shea, K., Krzywinski, M. et al. /// The SEIRS model for infectious disease dynamics /// *Nat Methods* 17, 557–558 (2020). https://doi.org/10.1038/s41592-020-0856-2

Fisman et al. (2022): Fisman DN, Amoako A, Tuite AR /// Impact of population mixing between vaccinated and unvaccinated subpopulations on infectious disease dynamics: implications for SARS-CoV-2 transmission /// *CMAJ* 2022;194:E573-80. https://doi.org/10.1503/cmaj.212105.